\documentclass[twocolumn,aps,floats,letterpaper,floatfix,groupedaddress]{revtex4-1}
\usepackage{amsfonts}
\usepackage{amssymb}
\usepackage{bm}
\usepackage{dsfont}
\usepackage{graphicx}
\usepackage{epsfig,float,afterpage}
\usepackage[countmax]{subfloat}
\usepackage{mathtools}
\usepackage[colorlinks,linkcolor=blue,citecolor=blue,urlcolor=blue]{hyperref}
\usepackage{dcolumn}
\usepackage{epstopdf}
\usepackage{color}
\usepackage{float}
\usepackage{amsmath}
\usepackage{physics}

\usepackage{siunitx}
\usepackage{xcolor}
\usepackage{ulem}
\usepackage{soul}       

\begin{document}

\title{Detection of Spin Coherence in Cold Atoms via Faraday Rotation Fluctuations}
\author{Maheswar Swar}
\author{Dibyendu Roy}
\author{Subhajit Bhar}
\author{Sanjukta Roy}
\author{Saptarishi Chaudhuri}

\affiliation{Raman Research Institute, C. V. Raman Avenue, Sadashivanagar, Bangalore 560080, India} 

\begin{abstract}
We report non-invasive detection of spin coherence in a collection of Raman-driven cold atoms using dispersive Faraday rotation fluctuation measurements, which opens up new possibilities of probing spin correlations in quantum gases and other similar systems. We demonstrate five orders of magnitude enhancement of the measured signal strength than the traditional spin noise spectroscopy with thermal atoms in equilibrium. Our observations are in good agreement with the comprehensive theoretical modeling of the driven atoms at various temperatures. The extracted spin relaxation rate of cold rubidium atoms with atom number density $\sim$10$^9/$cm$^3$ is of the order of 2$\pi\times$0.5 kHz at 150 $\mu$K, two orders of magnitude less than $\sim$ 2$\pi\times$50 kHz of a thermal atomic vapor with atom number density $\sim$10$^{12}/$cm$^3$ at 373 K. 
\end{abstract}


\maketitle
The prospects of non-invasive measurement schemes have found increasing research interests in the recent decades to detect equilibrium and non-equilibrium properties of microscopic and mesoscopic quantum systems \cite{Luca2018, Hammerer2010}. The modern scientific disciplines, in particular, quantum information science \cite{Suter2016}, quantum sensing \cite{Degen2017} and metrology \cite{Mitchell2020} can take advantage of the direct applications of these non-destructive measurements. Optical measurements employing dispersive light-matter interactions such as detecting the Faraday or Kerr rotation of an off-resonant probe light are examples of such measurements that disturb the measured samples minimally. The non-destructive optical Faraday and Kerr rotation measurements have been proposed and applied to a broad range of systems, including the read-out of a single electron's spin state in a semiconductor quantum dot \cite{quantum_dots,Atatre2007, Mikkelsen2007} and the quantum gas microscope \cite{2017_Takahashi, Kristensen_2017, Gajdacz2013,Yang2018, Meineke2012} for site-resolved imaging of single isolated atoms in an optical lattice.

In the absence of finite magnetization along the light propagation direction, the dynamical magnetic properties of the sample can be found from the temporal fluctuations of dispersive Faraday rotation. Such Faraday rotation noises have been extensively studied within the spin noise spectroscopy (SNS) \cite{Zapasskii2013, Kozlov2021, Sinitsyn2016} technique to detect the intrinsic spin dynamics in atomic vapors \cite{Crooker2004, Swar2018, Lucivero2016, Tang2020}, semiconductor heterostructures \cite{Oestreich2005}, quantum dots \cite{Roy2013,Glazov2013QDs}, spin-exchange collisions \cite{Roy2015, Katz2015} and exciton-polaritons \cite{Glazov2013, Ryzhov2020}. The SNS is also applied for precision magnetometry by using a spectral resolution of the spin noise signals from thermal atomic vapors \cite{Crooker2004, Swar2018} or semiconductors \cite{Ryzhov2016}.     

However, Faraday rotation fluctuation signals have not been detected so far in ultracold atoms and quantum gases, where direct measurement of spin fluctuations is predicted to be extremely useful in understanding quantum phases \cite{Recati2011, Altman2004, Roy2015b, Jepsen2020} and precision magnetometry \cite{Cohen2019, Gosar2021}. Moreover, the cold atomic systems are ideal testbeds for demonstrating quantum effects due to their ultra-low temperatures \cite{Bloch2008}.

 In this Letter, we focus on detecting spin coherences in cold atoms using Faraday-rotation fluctuation measurements. We theoretically develop and experimentally realize this measurement method to demonstrate an enhancement of signal strength as much as 10$^5$ in thermal atoms, which allows us to detect spin coherence in ultracold atoms. 

\begin{figure}[ht]
\centering
\includegraphics[scale=0.255]{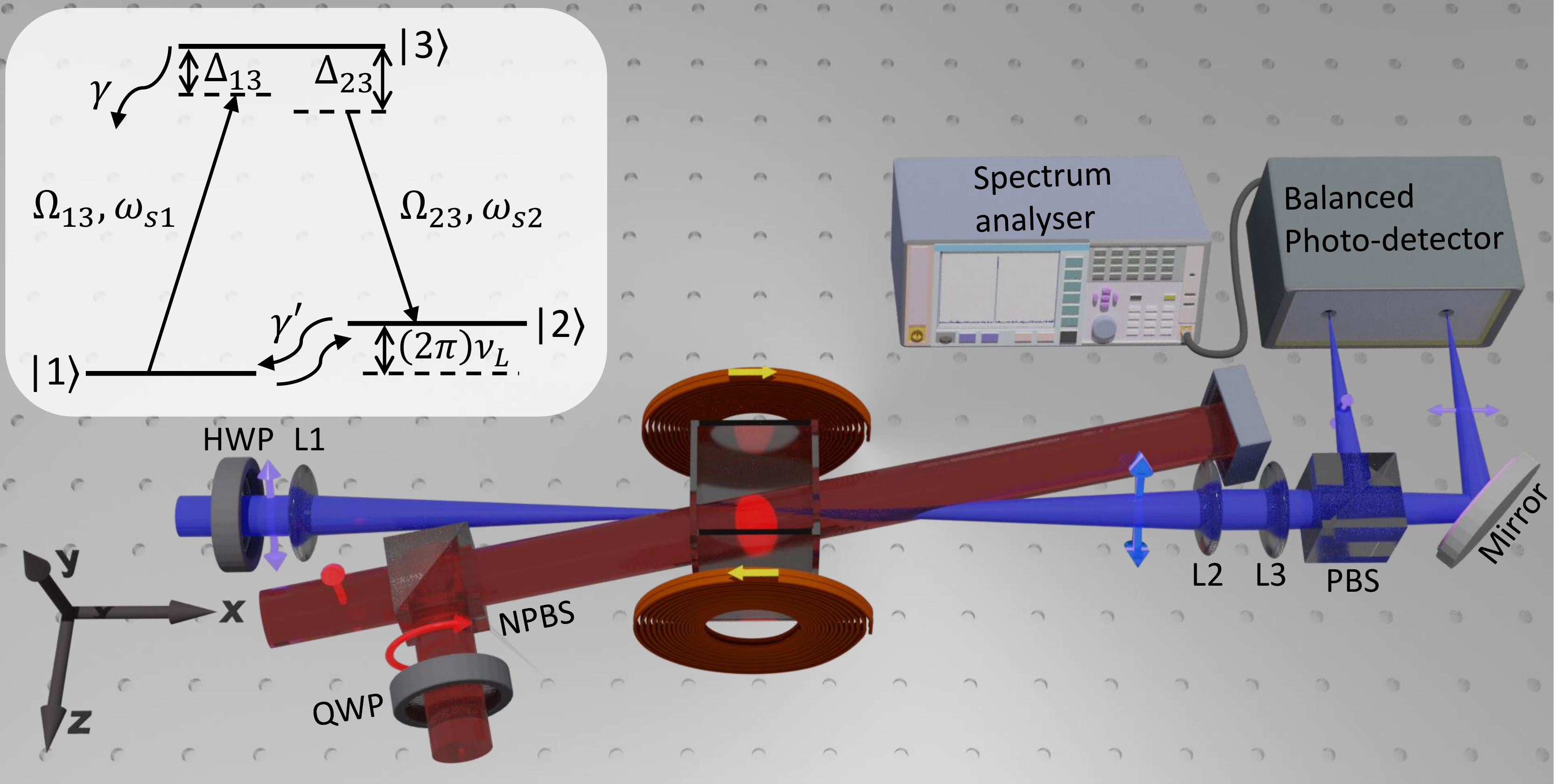}
\caption{The schematic of the experiment showing the Raman fields (red), probe laser (blue), the atomic cloud (red blob), magnetic coils, and detection set-up. (Inset) Level diagram used for coherent coupling between the Zeeman states ($|1\rangle$ and $|2\rangle$). $\Delta_{i3}, \Omega_{i3}$, and $\omega_{si}$ ($i =$ 1,2) are the optical detuning, Rabi angular frequency, and angular frequency of two Raman fields. $\gamma$ is the excited state ($|3\rangle$) linewidth, and $\gamma'$ is the spin relaxation rate between the states $|1\rangle$ and $|2\rangle$. The Raman resonance condition is satisfied when the frequency difference between the Raman fields coincides with the frequency difference between the states $|1\rangle$ and $|2\rangle$ (i.e., $\omega_{s1} - \omega_{s2}$ = $2\pi \nu_L$).}
\label{fig:fig2}
\end{figure}

A pair of phase-coherent Raman radiation fields derived from an external cavity diode laser (ECDL) interacts with a $\Lambda$-type three-level system (3LS) formed by two ground states $|1\rangle$, $|2\rangle$ and one excited state $|3\rangle$ as depicted in Fig.~\ref{fig:fig2} and its inset. The frequency, intensity, and polarization of the Raman fields are controlled by separate acousto-optic modulators (AOMs) and wave-plates. 

The semiclassical interaction Hamiltonian of the 3LS with the Raman fields can be written as \cite{scully1997quantum}:
\begin{eqnarray}
\begin{split}
\frac{\mathcal{H}}{\hbar}&= (\Delta_{23}-\Delta_{13}) \mu^{\dagger}\mu - \Delta_{13}\sigma^{\dagger}\sigma - \Omega_{13}(\sigma + \sigma^{\dagger})\\
&- \Omega_{23}(\mu + \mu^{\dagger}),
~\label{rad}
\end{split}
\end{eqnarray}
where, we define the dipole transition operators of the 3LS by $\sigma^{\dagger} =|1\rangle\langle3|$, $\mu^{\dagger} =|2\rangle\langle3|$, $\nu^{\dagger} =|1\rangle\langle2|$. Here, $\Omega_{13}$ and $\Omega_{23}$ are the resonant Rabi angular frequency of the Raman field 1 (R1) and Raman field 2 (R2), respectively. The detunings of the Raman fields from the related optical transitions are $\Delta_{13} = \omega_{s1}-\omega_3 + \omega_1$, $\Delta_{23} = \omega_{s2}-\omega_3 + \omega_2$, where $\omega_{s1} (\omega_{s2})$ is the angular frequency of the Raman field 1(2), and $\omega_i$ is that of the state $|i\rangle~(i=1,2,3)$. The symbols $s_1, s_2$ denote the polarizations (linear or circular) of the two Raman fields.

An off-resonant, linearly polarized probe laser field generated from another ECDL propagating along $x$-direction dispersively detects the temporal fluctuations of the population between the Zeeman states within a ground hyperfine level of rubidium atoms. The measured temporal Faraday rotation fluctuations of the probe field are proportional to population fluctuations represented by a two-time correlation $\langle \tilde{\rho}_{21}^{\dagger}(t)\tilde{\rho}_{21}(0)\rangle$ of density matrix coherence $(\tilde{\rho}_{21})$ between these states in the laboratory frame. We write the power spectrum by taking the Fourier transform of such a correlation:
\begin{eqnarray}
P(\omega)&=&\frac{1}{2\pi}\int_{-\infty}^{\infty}dt\: e^{i\omega t} \langle \tilde{\rho}_{21}^{\dagger}(t)\tilde{\rho}_{21}(0)\rangle,
\end{eqnarray}
where the expectation is performed over equilibrium thermal noise.

In the absence of driving ($\Omega = 0$) by the Raman fields, the intrinsic spin noise (SN) power spectrum for the spontaneous fluctuations of the population in equilibrium is given by \cite{MS2021}
\begin{eqnarray}
P(\omega)|_{\Omega=0}=\frac{N_2}{2\pi}\frac{\gamma_{21}}{\gamma_{21}^2+(\omega-\omega_2+\omega_1)^2}, \label{inPS}
\end{eqnarray}
where  $\gamma_{ij}$ is the relaxation rate from state $i$ to $j$ of the atoms and $N_2$ is the number of atoms in the observation region. $P(\omega)|_{\Omega=0}$ is a Lorentzian centered around $\omega = (\omega_2-\omega_1) \coloneqq 2\pi \nu_L$ and with a full-width at half maximum (FWHM) of $2\gamma_{21}$. 

In the presence of the Raman fields, the atoms are driven out of equilibrium, and the population of the Zeeman states starts to oscillate coherently with increasing $\Omega_{13}$ and $\Omega_{23}$. For a strong driving, the intrinsic fluctuations (related to equilibrium noise) form a broad background in the measured power spectrum. Therefore, we ignore the equilibrium noise in leading order to obtain a simple expression for the power spectrum of the strongly driven atoms at steady-state \cite{MS2021}: 
\begin{eqnarray}
\begin{split}
P(\omega) =\delta(\omega + \omega_{s2} - \omega_{s1})\vert \rho_{21} \vert ^2,
\label{pwrspec}
\end{split}
\end{eqnarray}
where $\tilde{\rho}_{21}(t)=\rho_{21}\:e^{i(\omega_{s_2}-\omega_{s_1})t}$. We can find a relatively simple formula for $\rho_{21}$ by choosing $\Omega_{13}=\Omega_{23}=\Omega,\gamma_{13}=\gamma_{23}=\gamma,\gamma_{12}=\gamma_{21}=\gamma'$ and $\Delta_{23}=0,\Delta_{13}=\Delta$:
\begin{widetext}
\begin{eqnarray}
\rho_{21}&=& -\frac{\gamma \Omega^2 ((\gamma'+i\Delta)(2\gamma+i\Delta)+4\Omega^2)}{\gamma(\gamma'^2+\Delta^2)(2\gamma^2+\Delta^2)+(2\gamma\gamma'(3\gamma'+4\gamma)+(\gamma'+2\gamma)\Delta^2)\Omega^2+4(3\gamma'+2\gamma)\Omega^4}.
\label{coh}
\end{eqnarray}
\end{widetext} 
$P(\omega)$ in Eq.~\ref{pwrspec} shows a delta peak at $\omega = \omega_{s_1}-\omega_{s_2} \coloneqq 2\pi \delta_{12}$, whose strength $|\rho_{21}|^2$ in Eq.~\ref{coh} grows rapidly with increasing $\Omega$ before saturating at large $\Omega$. 

We first demonstrate the signal-strength enhancement with the Raman driving than the intrinsic SN signal (i.e., $\Omega=0$) in thermal atomic vapor. A vapor cell containing $^{87}$Rb atoms and neon buffer gas is placed in a uniform magnetic field (along $\hat{z}$) produced by a Helmholtz coil. We select here $|F = 2, m_{F} = -1\rangle \equiv |1\rangle$, $|F = 2, m_{F} = 0\rangle \equiv |2\rangle$ and $|F' = 3, m_{F'} = 0\rangle \equiv |3\rangle$. We align the Raman fields with a small angle $\sim 3^\circ$ to the $x$-axis to avoid them falling on the photo-detector.

\begin{figure}
\centering
\includegraphics[scale=0.52,trim=0 0 0 0, clip]{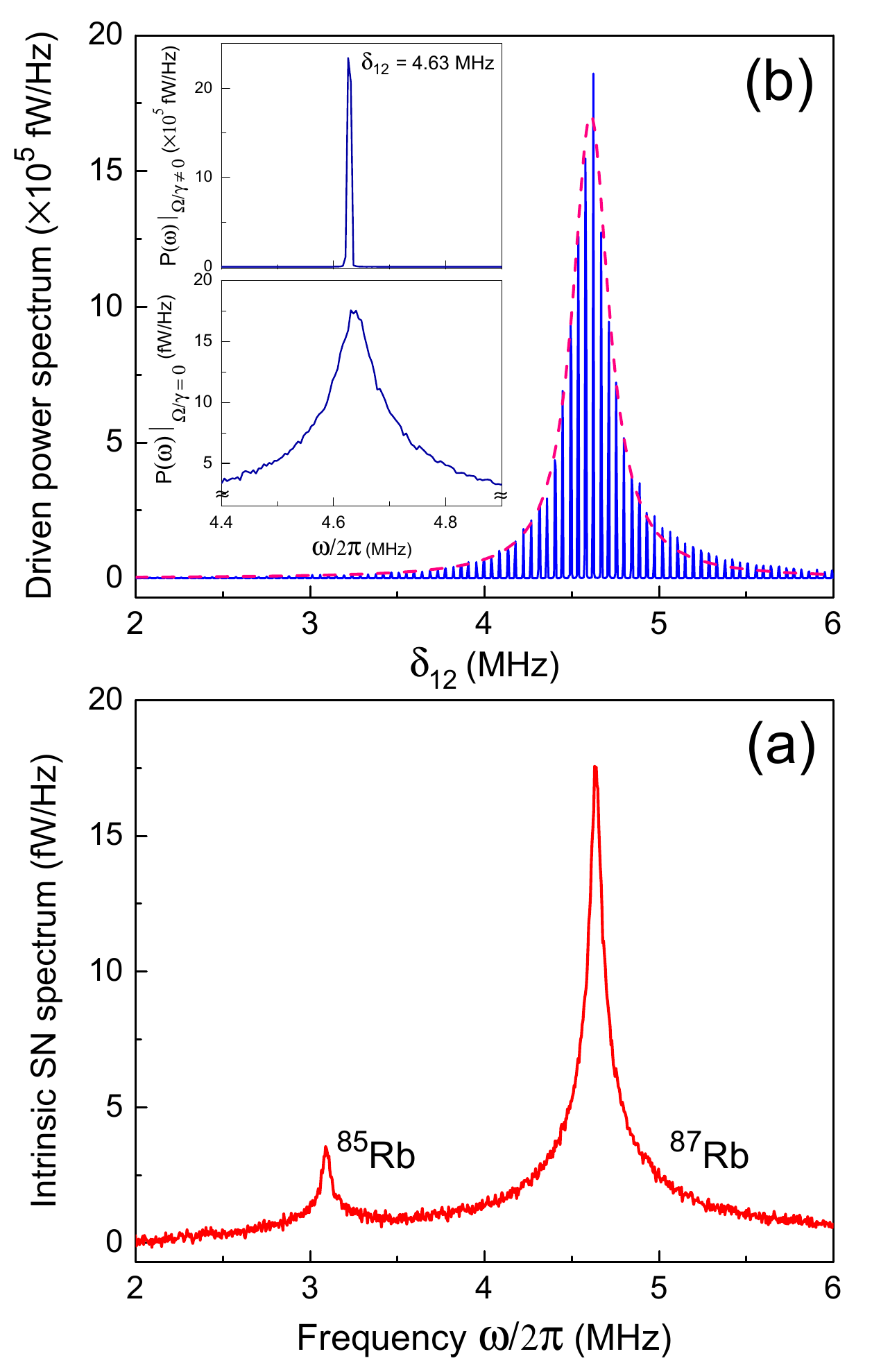}
\caption{Intrinsic spin noise (SN) spectrum (a) and Raman driven power spectrum (b) of rubidium atomic vapor at  T = 388 K. In (b), $^{87}$Rb atoms were coherently driven by a pair of Raman fields. The strength of the driven spectrum is enhanced by approximately 10$^5$ times in comparison to the intrinsic SN signal when the Raman field intensity $\Omega/\gamma = $7.6 $\times$ 10$^{-3}$. The insets in the top panel show a comparison between the driven spectrum at Raman resonance (top) and the intrinsic spectrum (bottom).}
\label{fig:fig3}
\end{figure}

The probe laser field is detuned by $- 4\gamma$ \footnote{The pressure broadened line-width, $\gamma = 2\pi \times$ 1.79(0.02) GHz, measured by absorption spectroscopy} from $5S_{1/2}, F = 2$ $\rightarrow$ $5P_{3/2},F' = 3$ (D$_2$) transition of $^{87}$Rb. The measured $P(\omega)|_{\Omega=0}$ is presented in Fig.~\ref{fig:fig3}(a), which shows the Lorentzian FWHM of $\sim$ 150 kHz for $^{87}$Rb atoms in thermal equilibrium at T = 388 K. The details of the instrumentation used can be found in \cite{Swar2018, Mugundhan2020}.

In presence of the Raman fields ($\Omega \neq 0$), we observe an enhancement of $P(\omega)$, as shown in Fig.~\ref{fig:fig3}(b) \footnote{Experimentally we observed a small but finite width ($\sim$ 2 kHz) of the individual spectrum instead of the delta function as predicted in Eq.~\ref{pwrspec}. This width is due to the relative frequency jitter of the two Raman fields derived from two independent AOMs}. The peak height is maximum when $\Delta=0$ (i.e., $\delta_{12} = \nu_{L}$), and it decreases with increasing $\Delta$. The envelope of the narrow peaks with varying $\delta_{12}$, as shown in Fig.~\ref{fig:fig3}(b), has a Lorentzian form when $\Omega<\gamma$.  But it switches to a Gaussian shape for larger $\Omega$. 

In the insets of Fig.~\ref{fig:fig3}(b), we show a comparison between the spectrum observed in the Raman-resonance condition for $\Omega/\gamma = 7.6\times10^{-3}$ (top) and that at $\Omega/\gamma = 0$ (bottom). An enhancement by 10$^5$ in the signal strength ($|\rho_{21}|^2$) has been observed for the driven system. In the subsequent experiments, we vary $\delta_{12}$ while keeping $\Delta_{23}=0$, and record a series of spectra shown in Fig.~\ref{fig:fig3}(b). These series of spectra span an envelope that is fitted by Eq.~\ref{coh} (pink dashed line), which gives a peak position at $\nu_{L}$ and an envelope width of $\sim 250$ kHz. 

\begin{figure}[ht]
\centering
\includegraphics[scale=0.30]{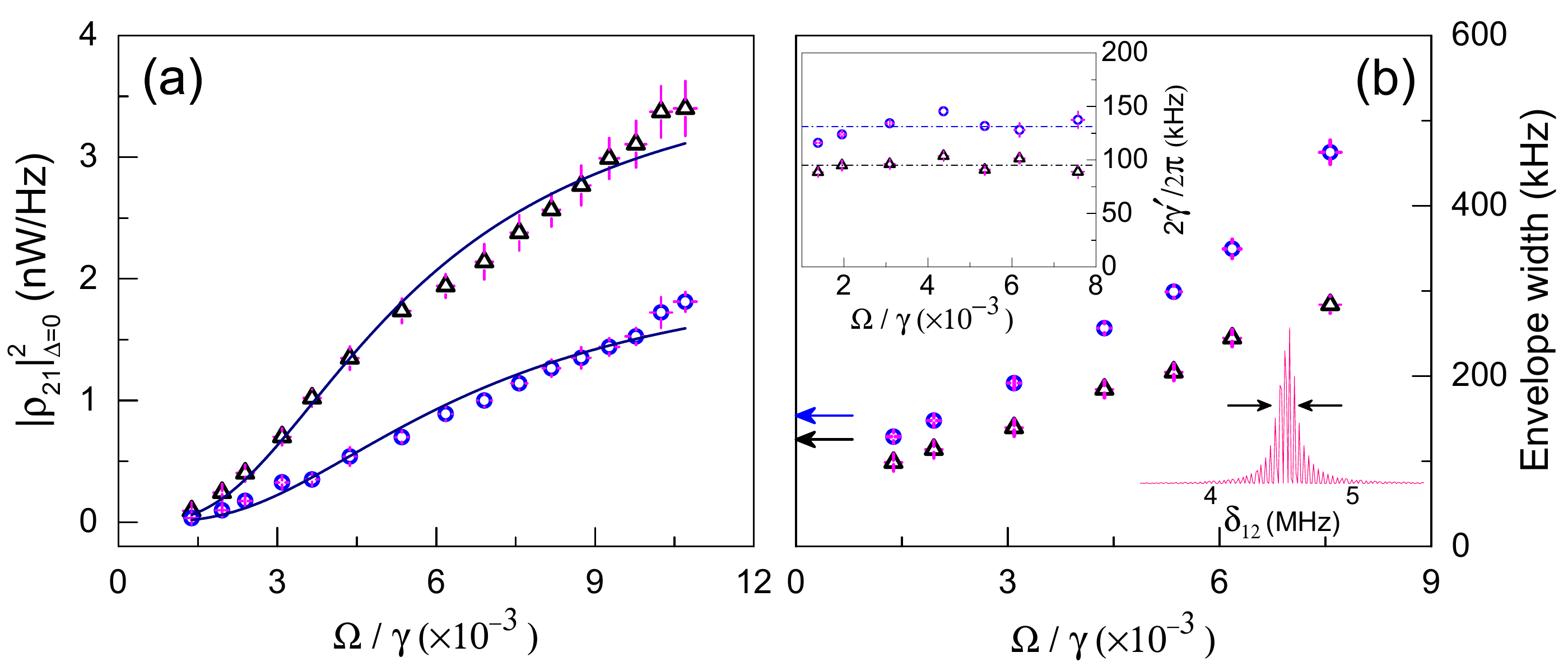}
\caption{On-resonance signal strength (a) and FWHM of the envelope of $P(\omega)$ (b) from the thermal vapor of $^{87}$Rb as a function of $\Omega/\gamma$. (a) The black triangles (blue circles) are the on-resonance peak strength for temperature T = 373 K (393 K). The solid lines are the fits by Eq.~\ref{Str}. (b) The black triangles (blue circles) represent the FWHM of the envelope fitted with Lorentzian profile for T = 373 K (393 K). In the inset, the black triangles (blue circles) show the extracted value of 2$\gamma'$ after fitting the envelope using Eq.~\ref{coh}. The black (blue) arrow indicated in the bottom-left side represents the measured FWHM of the intrinsic SN spectrum. The raw spectrum and its measured FWHM are indicated at the bottom-right corner.}
\label{fig:fig5}
\end{figure}

\begin{figure*}[ht]
\centering
\includegraphics[scale=0.52, trim =0cm 0 1cm 0, clip]{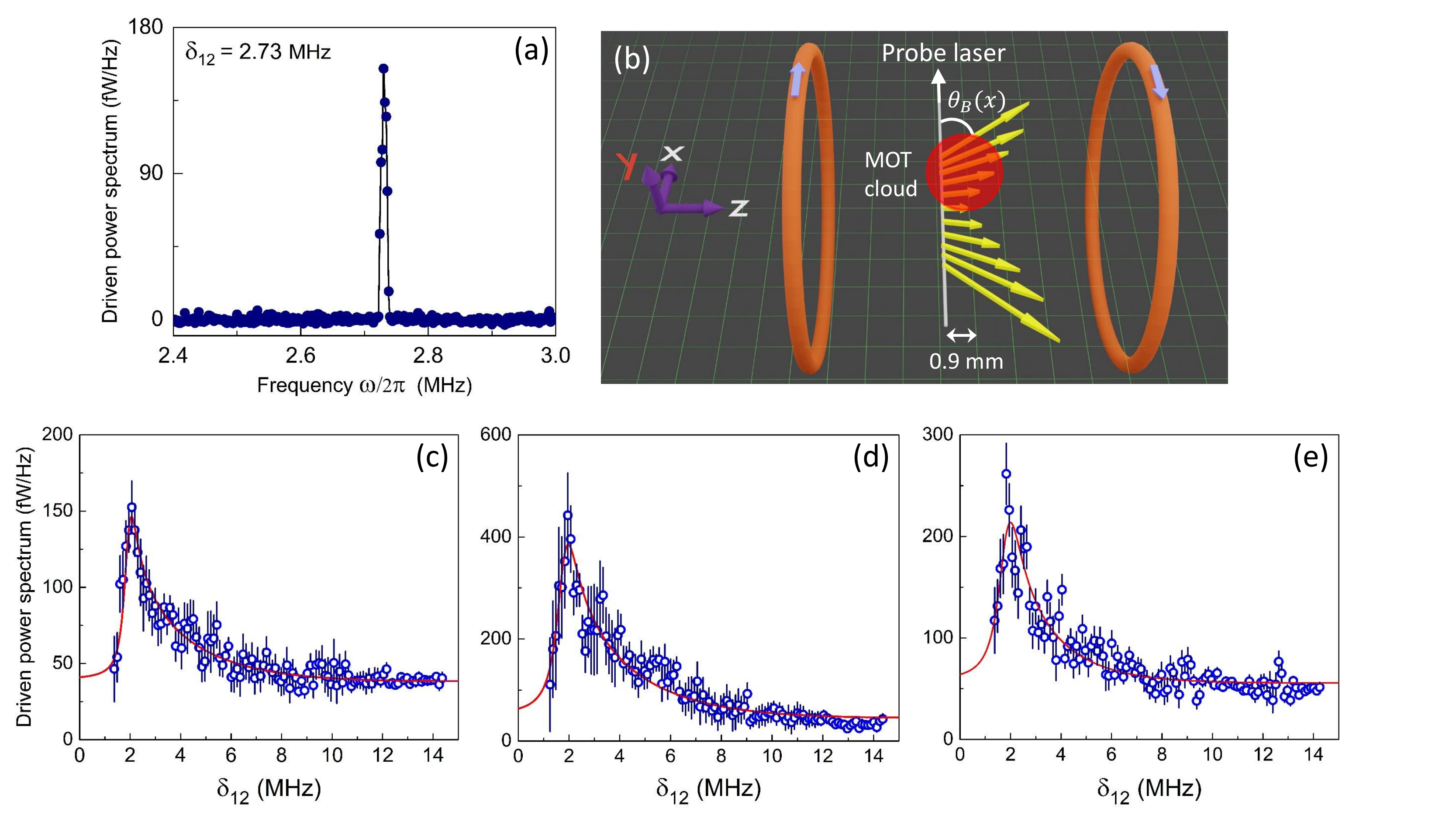}
\caption{Measurement of spin coherence in Raman driven cold $^{85}$Rb atoms at 150 $\mu$K. (a) $P(\omega)$ recorded at $\delta_{12} = 2.73$ MHz. (b) A schematic representation of our experiment including local magnetic field profiles (yellow arrows), the position of cold atomic cloud (red blob), and probe laser. (c-e) The recorded envelope spectra (blue open circles) for various sets (check the text for details) of experimental parameters.}
\label{fig:fig6}
\end{figure*}

The dependence of $|\rho_{21}|^2$ on $\Omega$ can be understood by taking the Raman resonant limit $\Delta=0$ ($\delta_{12} = \nu_{L}$) in Eq.~\ref{coh}, and we get
\begin{eqnarray}
|\rho_{21}|_{\Delta=0}^2=\frac{\gamma^2\Omega^4}{(\gamma'\gamma^2+(3\gamma'+2\gamma)\Omega^2)^2},
\label{Str}
\end{eqnarray}
which shows that the coherence between the ground levels grows with increasing $\Omega$ before saturating for higher $\Omega>\gamma$. In order to measure $|\rho_{21}|_{\Delta=0}^2$, we vary the intensities of the Raman fields keeping $\Delta=0$. The polarization of the Raman fields are $(\pi_{1})_x - (\sigma_{2}^{+})_x$ for the measurements presented in Fig.~\ref{fig:fig3} (b) and Fig.~\ref{fig:fig5}, where the subscript 1(2) refers to R1 (R2), and $x$-axis is the propagation direction of the Raman fields \cite{MS2021}. The measured on-resonance peak strength $|\rho_{21}|_{\Delta=0}^2$ as a function of $\Omega/\gamma$ is plotted in Fig.~\ref{fig:fig5}(a). The black triangles (blue circles) are the data corresponds to T = 373 K (393 K) of the cell. We fit these data by Eq.~\ref{Str} (solid lines) keeping only $\gamma'$ as a free parameter. We extract the value of $2\gamma'$ to be 2$\pi\times$(95$\pm$7) kHz and 2$\pi\times$(136$\pm$15) kHz for 373 K and 393 K, respectively. We have separately measured the FWHM of the intrinsic SN spectrum to be 2$\pi\times$(126$\pm$3) kHz and 2$\pi\times$(153.7$\pm$0.4) kHz for these temperatures, respectively. We attribute these small but finite (within 25 $\%$) differences to two competing effects of different physical origins - the perturbation induced by the Raman driving to bring the atoms beyond thermal equilibrium and linear response, and the suppression of spin projection noise due to coherent coupling.  

For a fixed $\Omega$, we vary $\delta_{12}$ and record the envelope of $P(\omega)$. We repeat these experiments for various values of $\Omega/\gamma$ and fit each spectrum with a Lorentzian function. The extracted FWHMs of the envelope for various $\Omega/\gamma$ are shown in Fig.~\ref{fig:fig5}(b). Black triangles (blue circles), shown in inset, are the extracted value of 2$\gamma'$ after fitting the envelope with $|\rho_{21}|^2$ given in Eq.~\ref{coh} for T = 373 K (393 K), respectively. The average value of $2\gamma'$, extracted from these measurements are 2$\pi\times$(95$\pm$6) kHz and 2$\pi\times$(131$\pm$10) kHz for those two temperatures, respectively. In the measurements reported in Fig.~\ref{fig:fig5}(b) (inset), we notice consistently lower values of extracted 2$\gamma'$ than the intrinsic measurements ($\Omega = 0$), which indicates the spin projection noise suppression is more significant for these thermal vapors than the perturbation effects bringing the system beyond equilibrium.

We have further experimentally verified that the envelope FWHM can be smaller than the intrinsic width of SN spectrum (indicated by arrow at the bottom-left side of Fig.~\ref{fig:fig5}(b)). While this occurs only for $\gamma'/\gamma>1$ when $\Omega/\gamma>1$, it can also occur for $\gamma'/\gamma<1$ when $\Omega/\gamma<1$. We have detected as much as 15$\%$ reduction in the width of driven envelope than the intrinsic SN spectrum.

We next implement the coherent Raman drive technique in a magneto-optically trapped (MOT) cold $^{85}$Rb atomic cloud to extract the spin relaxation rate $\gamma'$. To the best of our knowledge, this is the first detection of spin fluctuations in cold atoms using minimally invasive Faraday rotation fluctuation measurement. Our experimental system typically traps more than 10$^7$ atoms at a temperature of 150$\pm$10 $\mu$K in a standard vapor loaded MOT with a typical Gaussian width of $\sim$4 mm. We take sufficient care to ensure that the center of the atomic cloud is overlapped with the zero of the quadrupole trapping field within 30 $\mu$m. More details about the experimental system are provided in \cite{MS2021}. Additionally, we align a pair of Raman fields with waist diameter of 6 mm, blue detuned by 2$\gamma$ ($\gamma = 2\pi\times$6.1 MHz) from $5S_{1/2}, F = 3$ $\rightarrow$ $5P_{3/2},F' = 4$ (D$_2$) transition. A collimated probe laser field along $\hat{x}$ with a waist diameter 70 $\mu$m and blue detuned by 20$\gamma$ is sent through the cold atomic cloud (see Fig.~\ref{fig:fig2}). The probe beam position from the cloud center is independently measured using high precision absorption spectroscopy and absorption imaging within a precision of 10 $\mu$m.  The quadrupole magnetic field profile creating the MOT is separately characterized using a Hall magnetometer probe, and it is shown schematically in Fig.~\ref{fig:fig6}(b). 

To measure $P(\omega)$ at a finite Larmor frequency in the presence of the Raman fields, we have shifted the probe laser by 900 $\mu$m in $z$-direction on x-z plane from the trap center. Also, due to the Raman beams, the cloud center shifts in $x$-direction depending on the Raman beam intensity and the detuning (see Fig.~\ref{fig:fig6}(b)). In Fig.~\ref{fig:fig6}(a), we show a representative spectrum detected in this condition at $\delta_{12} = 2.73$ MHz for $\Omega/\gamma = 0.35$. In the absence of Raman fields - unlike in the vapor case -  we do not observe any detectable intrinsic SN signal due to the fact that the total number of atoms within the probe field is  10$^4$ times less than that of thermal vapor around 373 K. 

We record a series of $P(\omega)$ by varying $\delta_{12}$. For a fixed value of $\Omega/\gamma$ and probe beam position, we obtain a composite spectra (see Fig.~\ref{fig:fig6}(c)) by repeating the above procedure and recording the peak height of each individual $P(\omega)$. We analyze our experimental results using the earlier modeling for thermal vapors after including corrections due to magnetic field variations, atom density distributions, and multilevel contributions \cite{MS2021}.  We fit the data in Fig.~\ref{fig:fig6}(c) (solid red line) with a free parameter $\gamma'$. We get an estimate for the value of 2$\gamma'$ from this fitting as 2$\gamma'$ = 2$\pi\times$(1.0 $\pm$0.7) kHz.

We have repeated the above measurements for various $z$-position of the probe laser. Another representative data is shown in Fig.~\ref{fig:fig6}(d) for $z$ = 700 $\mu$m, which gives  2$\gamma'$ = 2$\pi\times$(2.3 $\pm$0.9) kHz. We have also performed the experiment in the presence of balanced Raman fields, which minimize the shift of the cloud center as shown in Fig.~\ref{fig:fig6}(e).  The extracted value of 2$\gamma'$ is 2$\pi\times$(1.3 $\pm$0.6) kHz. Our present experiment is limited by several factors, e.g.,  magnetic field inhomogeneity, off-resonant scattering, the effect of strong driving fields, and relative frequency stability of the Raman fields, which can significantly change the measured spin relaxation rate. At this low temperature (150 $\mu$K), the above perturbation effects are typically far more important than the suppression of spin projection noise discussed earlier. Nevertheless, the observed reduction in extracted spin relaxation rate by two orders of magnitude than the thermal atoms is in expected lines, which can be attributed to the six-order lowering of the temperature that substantially reduces thermal coupling, collisions, and transit times.

In conclusion, we report relatively non-invasive detection of spin coherence in cold atoms driven coherently by Raman fields. Such detection method can have significant potential applications for precision magnetometry, high-resolution imaging, non-perturbative probing of quantum phase transitions in cold atoms and other similar systems, e.g., cold ions, cold molecules, etc.

This work was partially supported by the Ministry of Electronics and Information Technology (MeitY), Government of India, through the Center for Excellence in Quantum Technology, under Grant4(7)/2020-ITEA. S. R acknowledges
funding from the Department of Science and Technology, India, via the WOS-A project grant no. SR/WOS-A/PM-59/2019. We acknowledge Hema Ramachandran, Fabien Bretenaker, Priyanka G.L., Meena M.S., Sagar Sutradhar, Snehal Dalvi and RRI workshop for instruments, discussions and technical assistance.

\bibliographystyle{apsrev4-1} 
\bibliography{ColdSNSBib}

\end{document}


\title{Supplemental material for ``Detection of Spin Coherence in Cold Atoms via Faraday Rotation Fluctuations"}
\author{Maheswar Swar}
\author{Dibyendu Roy}
\author{Subhajit Bhar}
\author{Sanjukta Roy}
\author{Saptarishi Chaudhuri}

\affiliation{Raman Research Institute, C. V. Raman Avenue, Sadashivanagar, Bangalore 560080,  India}

\maketitle
\setcounter{figure}{0}
\renewcommand\thefigure{S\arabic{figure}}
\setcounter{equation}{0}
\renewcommand\theequation{S\arabic{equation}}

 This supplemental material provides details of our experimental set-up, theoretical modeling, and data analysis methods. We also discuss the dependence of the Faraday rotation fluctuations signal on the polarization states of the Raman fields.

\subsection{Description of the experimental set-up} 

In the present Letter, we have reported the results from two sets of experiments: one in a thermal atomic vapor to develop and characterize the measurement of Faraday rotation fluctuations signal from a coherently driven system, and another to perform the main study of the extraction of spin relaxation rate from a cloud of cold $^{85}$Rb atoms. Here, we provide details of the second set-up. We magneto-optically trap neutral $^{85}$Rb atoms inside a glass cell maintaining background pressure less than $10^{-10}$ millibar using standard laser cooling and trapping techniques. The cooling beams were generated from an external cavity diode laser (ECDL) and frequency stabilized to the 12 MHz red detuned with respect to $5S_{1/2}, F = 3$ $\rightarrow$ $5P_{3/2},F' = 4$ (D$_2$) transition. The `repumping' laser beams were derived from another ECDL and frequency stabilized to the $5S_{1/2}, F = 2$ $\rightarrow$ $5P_{1/2},F' = 3$ (D$_1$) transition. A pair of magnetic coils in near ideal anti-Helmholtz configuration produces the required spatial magnetic field gradient. We coincided the optical and magnetic field centers with the center of the vacuum chamber with an accuracy of $\sim$ 30 $\mu$m. We use three independent detection techniques: absorption imaging, fluorescence imaging, and probe absorption for characterizing the cold atoms and alignment of the optical fields for the experiments.

\begin{figure}[h]
\centering
\includegraphics[scale=0.45,trim=0 0 0 0, clip]{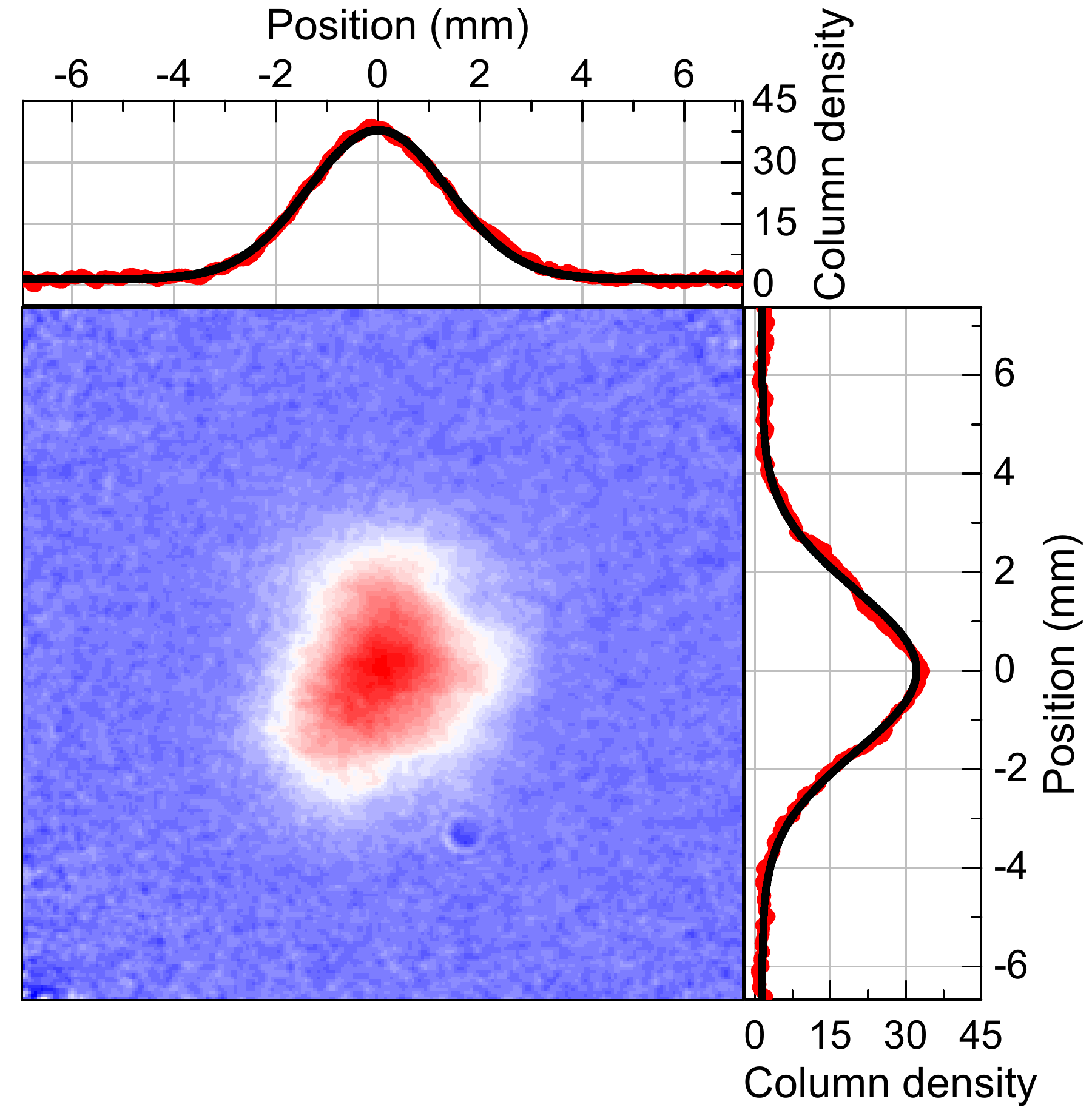}
\renewcommand{\thefigure}{S\arabic{figure}}
\caption{The absorption image of magneto-optically trapped (MOT) cold $^{85}$Rb atomic cloud. The top and right insets show the column density profiles (red) and fit to Gaussian (black) of the trapped cloud.}
\label{fig:s1}
\end{figure}

In Fig.~\ref{fig:s1}, we show a typical absorption image of our cold cloud, which gives a good measurement of the total number and the spatial distribution of the atoms. Using a separate time-of-flight measurement, we obtain the temperature of the atomic cloud to be $\sim$ 150 $\mu$K. We typically trap more than 10$^7$ atoms with a Gaussian full width at half maxima (FWHM) of $\sim$ 4 mm. 

 The probe laser field is generated from yet another ECDL, and its frequency is monitored using a high precision wavelength meter (Highfinesse, WSU2) with an absolute frequency accuracy of 1 MHz. The collimated probe beam with a Gaussian waist diameter of 70 $\mu$m was sent through the atomic cloud along $\hat{x}$ at $y = 0$, and the $z$ position of the probe beam was varied for different sets of measurements. 

\begin{figure}[h]
\centering
\includegraphics[scale=0.40,trim=0 0 0 0, clip]{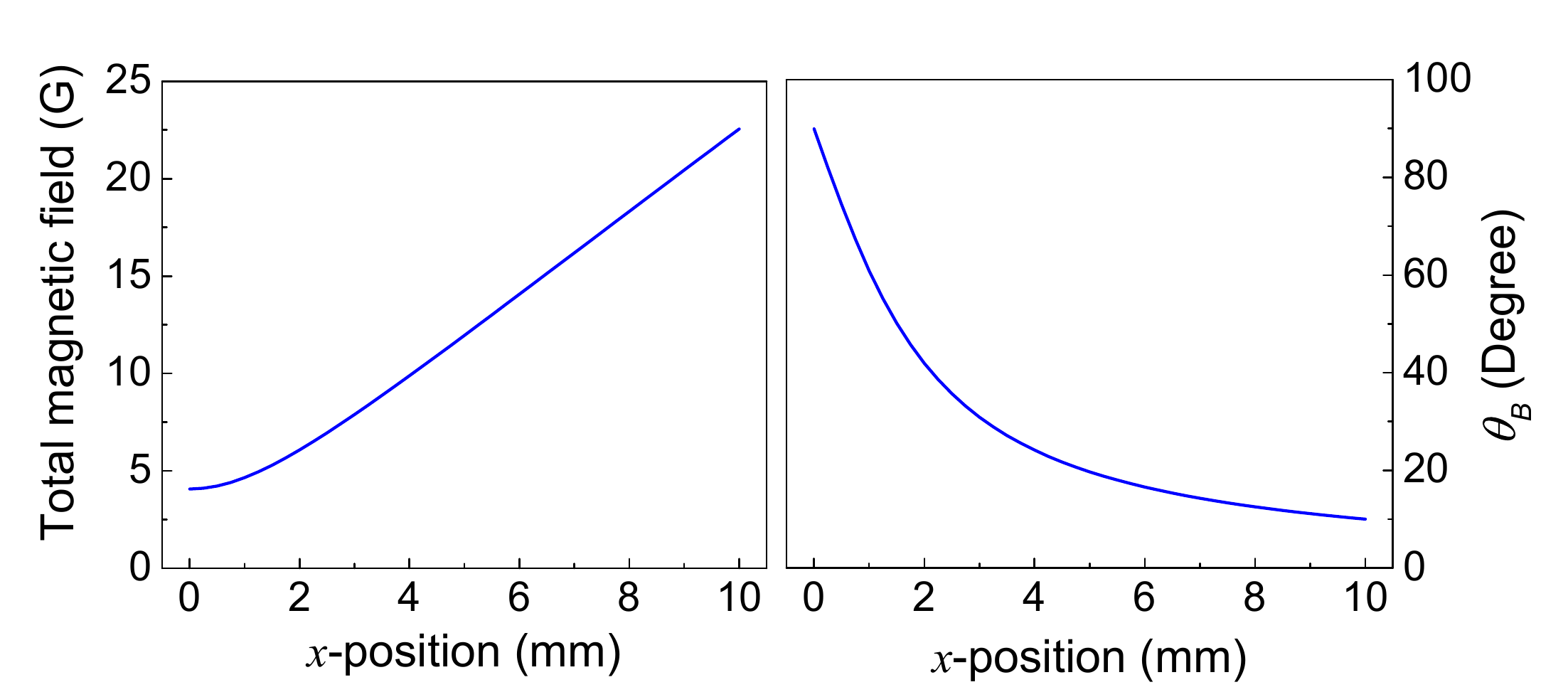}
\renewcommand{\thefigure}{S\arabic{figure}}
\caption{The $x$-dependence of the total magnetic field and the angle $\theta_B(x)$ between the local magnetic field and the probe laser propagation at $y = 0$ and $z = 900~\mu$m.}
\label{fig:s2}
\end{figure}

 The strength and orientation of the magnetic field $B$ vary within the atomic cloud along the probe field direction ($x$-axis) as shown in Fig.~\ref{fig:s2}. In this configuration, the magnetic field exists only on the x-z plane, and we calculate the magnetic field at each position using a solution of the elliptic equations \cite{Simpson}. We separately measure the field components ($B_x, B_y, B_z$) using a Hall probe magnetometer (LakeShore) for comparison and calibration purposes. The angle $\theta_B (x)$ between the local magnetic field and probe laser propagation direction is position-dependent.

The Raman radiation fields for coherent driving were derived from the same ECDL, which provides the cooling laser light. Two independent acousto-optic modulators (AOMs) were applied to produce the Raman fields with controllable frequency difference (denoted as $\delta_{12}$ in the main text). The radio frequency signal sent into the AOMs can be tuned using two voltage-controlled crystal oscillators. The control voltage was sent using an FPGA board with a vertical resolution of 12 bits. The long-term relative frequency `jitter' of $\delta_{12}$ was measured to be $\sim$ 7 kHz (for measurement duration of 8 seconds). This `jitter' limits our measurement precision, which we wish to reduce to a few Hz level using ultra-stable reference sources in our future upgrading of the experimental set-up. However, this relative frequency stability is adequate for the first sets of measurements of the Faraday rotation fluctuations signal from the cold atoms. The Raman fields were spatially mode cleaned using PM fibers, expanded to a Gaussian waist diameter of 6 mm, and combined in a non-polarizing cube beam splitter (NPBS) before sending through the cold atomic cloud. The polarization of the Raman fields can be independently varied, employing the combination of half-wave plates (HWPs) and quarter-wave plates (QWPs). We typically manage to obtain a polarization purity of the Raman fields $>$ 99$\%$.

\subsection{Details of the theoretical modeling}

{\it Intrinsic spin noise spectrum:} In the absence of driving by the Raman fields $(\Omega=0)$, the rubidium atoms are in thermal equilibrium, and the populations in their hyperfine ground level Zeeman states fluctuate over time due to thermal (and quantum) fluctuations. Such equilibrium population fluctuations generate an intrinsic spin noise (SN) in the atomic vapor. When $\Omega=0$, only the lower two levels of the three-level system (3LS) participate in the equilibrium spin dynamics detected by the probe beam. Therefore, we can derive the SN spectrum by including a noise term in the master equations of the density matrix elements of these levels. Thus, we write
\bea
\f{d\tilde{\rho}_{21}}{dt}&=&i(\om_2-\om_1+i\gamma_{21})\tilde{\rho}_{21}+\eta(t),
\eea
where we assume the noise $\eta(t)$ to be a Gaussian white noise with zero mean and $\la \eta(t') \eta(t'')\ra=N_{2}\gamma_{21}\delta(t'-t'')$, where $N_2$ is number of atoms within the measurement region. We then get the power spectrum $P(\om)|_{\Omega=0}$ of the spontaneous spin fluctuations as:
\bea
P(\om)|_{\Omega=0}&=&\f{1}{2\pi}\int_{-\infty}^{\infty}dt\: e^{i\omega t} \la \tilde{\rho}_{21}^{\dg}(t)\tilde{\rho}_{21}(0) \ra \nonumber\\&=&\f{1}{\pi}{\rm Re}\Big[\int_{0}^{\infty}dt\: e^{i\omega t} \la \tilde{\rho}_{21}^{\dg}(t)\tilde{\rho}_{21}(0)\ra\Big]\nn\\&=&\f{N_2}{2\pi}\f{\gamma_{21}}{\gamma_{21}^2+(\om-\om_2+\om_1)^2},
\eea
where $\omega$ is the spectral frequency.

{\it Raman-driven power spectrum:} In the presence of the Raman fields, the atoms are driven out of thermal equilibrium. We again write phenomenological master equations for the evolution of various components of the density matrix of the 3LS for the Hamiltonian $\mathcal{H}$ in the main text. Since the intrinsic fluctuations (related to the equilibrium noise, e.g., $\eta(t)$ above) only form a broad background in the measured power spectrum for a relatively strong driving by the Raman fields, we drop these noise terms from the following master equations to be able to extract an analytical expression for the measured power spectrum. These master equations are written in the rotating frame by rewriting the elements of coherence as $\rho_{31}(t)=\tilde{\rho}_{31}(t)e^{i\om_{s_1}t}, \rho_{32}(t)=\tilde{\rho}_{32}(t)e^{i\om_{s_2}t},\rho_{21}(t)=\tilde{\rho}_{21}(t)e^{i(\om_{s_1}-\om_{s_2})t}$, where $\tilde{\rho}_{31}(t),\tilde{\rho}_{32}(t)$ and $\tilde{\rho}_{21}(t)$ are the density matrix elements in the laboratory frame, and $\rho_{ij}$ are those in the rotated frame. We further take the following limits for the relaxation rates, $\gamma_{13}=\gamma_{23}=\gamma,\gamma_{12}=\gamma_{21}=\gamma', \gamma_{31}=\gamma_{32}=0$ and $\gamma' \ll \gamma$, to simplify the master equations.
\bea
\f{d\rho_{11}}{dt}&=&\gamma(1-\rho_{11}-\rho_{22})-i\Omega_{13}(\rho_{13}-\rho_{31}), \\
\f{d\rho_{22}}{dt}&=&\gamma(1-\rho_{11}-\rho_{22})-i\Omega_{23}(\rho_{23}-\rho_{32}), \\
\f{d\rho_{13}}{dt}&=&-(\gamma+i\Delta_{13})\rho_{13}-i\Omega_{23}\rho_{12}\nonumber \\&-&i\Omega_{13}(2\rho_{11}-1+\rho_{22}), \\
\f{d\rho_{31}}{dt}&=&-(\gamma-i\Delta_{13})\rho_{31}+i\Omega_{23}\rho_{21}\nonumber \\&+&i\Omega_{13}(2\rho_{11}-1+\rho_{22}), \\
\f{d\rho_{23}}{dt}&=&-(\gamma+i\Delta_{23})\rho_{23}-i\Omega_{13}\rho_{21}\nonumber \\&-&i\Omega_{23}(2\rho_{22}-1+\rho_{11}), \\
\f{d\rho_{32}}{dt}&=&-(\gamma-i\Delta_{23})\rho_{32}+i\Omega_{13}\rho_{12}\nonumber \\&+&i\Omega_{23}(2\rho_{22}-1+\rho_{11}), \\
\f{d\rho_{12}}{dt}&=&-i(\Delta_{13}-\Delta_{23}-i\gamma')\rho_{12}+i\Omega_{13}\rho_{32}\nonumber \\&-&i\Omega_{23}\rho_{13}, \\
\f{d\rho_{21}}{dt}&=&i(\Delta_{13}-\Delta_{23}+i\gamma')\rho_{21}-i\Omega_{13}\rho_{23}\nonumber \\&+&i\Omega_{23}\rho_{31}.
\eea
We apply these equations to investigate how the Raman fields affect the coherence $\tilde{\rho}_{21}(t)$ between the ground levels. From the above set of equations, we find $\rho_{21}(t)$  at the steady-state by setting $d\rho_{ij}(t)/dt=0$. Since we ignore the noise terms in the above master equations in the leading order of $\Omega_{13},\Omega_{23}$, we rewrite the power spectrum defined in the main text without the noise averaging:
\bea
P(\om)&=&\f{1}{2\pi}\int_{-\infty}^{\infty}dt\: e^{i\omega t} \tilde{\rho}_{21}^{\dg}(t)\tilde{\rho}_{21}(0)\nonumber \\
&=&\f{1}{2\pi}\int_{-\infty}^{\infty}dt\: e^{i(\omega-\om_{s_1}+\om_{s_2})t} \rho_{21}^{\dg}(t)\rho_{21}(0). 
\eea
When the driven atoms reach the steady-state at a long time, $\rho_{21}(t)$ becomes time-independent. Then, we can replace $\rho_{21}(t)$ and $\rho_{21}(0)$ by their steady-state value $\rho_{21}$ to find:
\bea
P(\om)&=&\f{1}{2\pi}\int_{-\infty}^{\infty}dt\: e^{i(\omega-\om_{s_1}+\om_{s_2})t} |\rho_{21}|^2\nonumber \\
&=&\delta(\omega+\om_{s_2}-\om_{s_1})|\rho_{21}|^2. \label{spectrum}
\eea
In the main text, we provide a relatively simple formula for $\rho_{21}$ by choosing $\Omega_{13}=\Omega_{23}=\Omega$ and $\Delta_{23}=0,\Delta_{13}=\Delta$, and that formula is employed for the driven thermal vapors. Here, we give a more general formula when $\Delta_{23}\ne 0$, which is our experimental condition for the driven rubidium cold atoms.
\begin{widetext}
\bea
\rho_{21}=\f{\gamma \Omega^2 (\tilde{\Delta}(2i\gamma+\tilde{\Delta})-4\Omega^2+(-2\gamma+i\tilde{\Delta})\gamma')}{\gamma(\tilde{\Delta}^2(2\gamma^2+\Delta_{23}^2+\Delta_{13}^2)+2\Omega^2\tilde{\Delta}^2+8\Omega^4)+\Omega^2(8\gamma^2+(\Delta_{23}+\Delta_{13})^2+12\Omega^2)\gamma'+\gamma \gamma'^2(2\gamma^2+\Delta_{23}^2+\Delta_{13}^2+6\Omega^2)},\nn\\\label{cohg}
\eea
\end{widetext}
where $\tilde{\Delta}=\Delta_{23}-\Delta_{13}$. The power spectrum in Eq.~\ref{spectrum} gives a delta peak (broadened in our experiment by relative frequency jitter of the two Raman fields derived from two independent AOMs) at $\omega=\om_{s_1}-\om_{s_2}$, whose strength is determined by $|\rho_{21}|^2$ given in Eq.~\ref{cohg}. The strength of the peak is maximum at $\Delta_{13}=0$ when $\Delta_{23}=0$, and the peak height falls with increasing $\Delta_{13}$. For $\Delta_{23}=0$, the envelope of the sharp delta peaks with changing $\Delta_{13}$ has a Lorentzian shape when $\Omega<\gamma$ but it changes to Gaussian form for larger $\Omega$. For $\Delta_{23} \ne 0$, the peak height is maximum at $\Delta_{13}=\Delta_{23}$. The envelope of $|\rho_{21}|^2$ is asymmetric with $\Delta_{13}$ when $\Delta_{23} \ne 0$.


\subsection{Simulation and data analysis}
 
For a homogeneous magnetic field as in the thermal vapor measurements, the Faraday rotation fluctuations signal (both the intrinsic and the Raman driven) is centered around a single Larmor frequency $\nu_L$ ($= g_F \mu_B B/h$, where $g_F$ is the Land$\acute{e}$ g-factor of the hyperfine F-levels, $\mu_B$ is the Bohr magneton and h is the Planck's constant), determined by the magnetic field strength $B$. However, for the cold atoms inside an MOT, the Larmor frequency $\nu_L(x)$ varies over space along $\hat{x}$. Therefore, the Zeeman splittings of the ground hyperfine levels are determined by the magnitude of the local magnetic field and, the Raman resonance condition is also position-dependent, i.e., $\delta_{12}(x) = \nu_L(x)$. In our numerical modeling, we take this position-dependent local magnetic field into account to calculate the strength of the Faraday rotation fluctuations signal at different frequency. 

Moreover, the orientation ($\theta_B(x)$) of the local magnetic field also varies along $\hat{x}$ inside the MOT. The SN signal strength at $\nu_L(x)$ also gets modified by a factor sin$^{2}\theta_B(x)$ \cite{Sinitsyn2016}. We have experimentally verified this correction factor by performing a separate calibration measurement of the intrinsic SN spectrum in thermal atoms.  

The atom density distribution within the MOT detected by the probe laser is not uniform, which is evident from the absorption image in Fig.~\ref{fig:s1}. We incorporate this density distribution in the modelling of the SN signal strength ($\propto n(x)$, where $n(x)$ is the number density of atoms at position $x$) from the MOT \cite{Crooker2004,MswarBook}. Note that $n(x)$ can be measured precisely using absorption imaging. 

Another minor correction to the SN signal strength can be from the definition of the quantization axis, which also varies along $x$-direction on the x-z plane. Since the local magnetic field alters over space, the coupling of the Raman fields with the atoms in the MOT also depends on $x$-position. Such correction can be incorporated in our modelling by an $x$-dependent Rabi frequency defined as $\Omega(x) = \Omega(1 - \sin^{2}\theta_B(x)/2)$.

So far, we have discussed the Raman-driven spin coherence between the ground states involving a single $\Lambda$ system formed by states $|i\rangle$, where $i$ = 1,2, and 3, as described in the paper. Ideally, in cold atom experiments, six $\Lambda$ systems are involved in giving rise to the driven power spectrum generated from $F =$ 3 and $F' =$ 4 hyperfine levels of $^{85}$Rb. However, the value of Land$\acute{e}$ g-factor is different for those two hyperfine levels, resulting in dissimilar contribution in building the signal strength from individual $\Lambda$ system through the optical detuning of the Raman fields.

Incorporating the above factors and corrections, we get the strength $|\rho_{21}(x)|^2$ of the Faraday rotation fluctuations signal from the atoms at position $x$ inside the MOT,
\begin{equation}
\vert \rho_{21}(x) \vert ^2 = 
\vert \rho_{21} (\Omega(x)) \vert ^2   n(x)  \sin^{2} \theta_B(x).
\label{Eqn:modrho}
\end{equation}

For a fixed position $x$ within the MOT, we consider the contributions of all six $\Lambda$ systems and add them up to get the total driven spectrum strength. The six $\Lambda$ systems contribute differently via the optical detuning $\Delta_{23}(n)$ of the Raman fields from the excited state, i.e.,
\begin{equation}
\Delta_{23}(n)= 2\gamma + (3 - n)(g_{F'} - g_F) \mu_B B (x)/h,
\label{Eqn:modrho1}
\end{equation}
 where $n$ runs from 1 to 6 and $g_{F'}$ is the Land$\acute{e}$ g-factor of the excited state. We fix the detuning $\Delta_{23}(n)= 2\gamma$ for $n=3$ in our experiment as described in the main text.
 
In Fig.~4(c,d,e) of the main paper, we present some plots for the Faraday rotation fluctuations signals as a function of $\delta_{12}$ from a driven cold rubidium cloud. We have fitted the experimental data using the Eq.~\ref{Eqn:modrho} along with $\rho_{21}(\Omega(x))$ from Eq.~\ref{cohg}, where $\Omega$ is replaced by $\Omega(x)$ and we employ $\Delta_{23}$ from Eq.~\ref{Eqn:modrho1}. The only free parameter in this fitting is the relaxation rate $\gamma'$ of the hyperfine ground level Zeeman states, and all other parameters are measured in our experiment.

\subsection{Dependence of driven power spectrum on Raman fields' polarization}
We have experimentally investigated the dependence of the Raman driven  power spectrum  on the polarization state of the Raman fields in both thermal vapors and cold atomic clouds. 
\par
\begin{figure}
\centering
\includegraphics[scale=0.35,trim=0cm 0 0 0, clip]{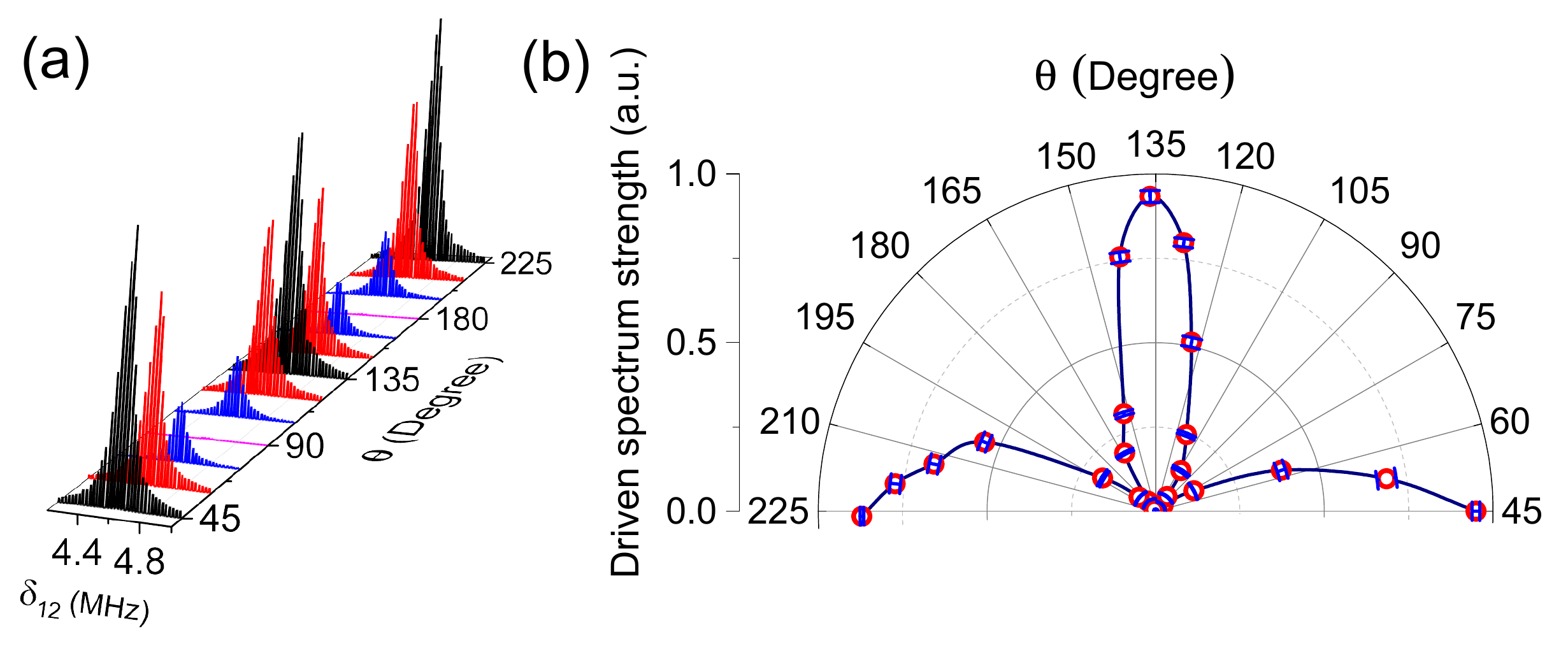}
\renewcommand{\thefigure}{S\arabic{figure}}
\caption{(a) The Raman driven power spectrum with detuning $\delta_{12}$ for various polarization states of the R2 field in thermal vapors. The R1 field is p-polarized ($(\pi_{1})_x$). $\theta$ is the angle between the optic axis of the quarter-wave plate (QWP) and the input polarization (p-polarized) of the R2 field. The angular momentum conservation of light-matter interactions in $\Lambda$ system is satisfied for $(\pi_{1})_x-(\sigma_{2}^{\pm})_x$ polarization combinations of the Raman fields. These correspond to $\theta = $ 45$^\circ$, 135$^\circ$, 225$^\circ$, where a maximum in the spectrum is observed. The driven spectrum vanishes for $(\pi_{1})_x-(\pi_{2})_x$ polarization combinations ($\theta =$ 90$^\circ$and 180$^\circ$), implying no coherent coupling between the states $|1\rangle \leftrightarrow |2\rangle$ by the Raman fields. (b) The driven spectrum strength as a function of $\theta$ at Raman resonance condition.}
\label{fig:s3}
\end{figure}

In Fig.~\ref{fig:s3} (a), we show the driven power spectrum for various combinations of the polarization state of the Raman fields in thermal vapors. The polarization of the R1 field is linear ($(\pi_{1})_x$), and kept fixed. We have tuned the polarization state of R2 field, and recorded the driven power spectrum as shown in Fig.~\ref{fig:s3}(a). The strength of the spectrum is maximum for $(\pi_{1})_x-(\sigma_{2}^{+})_x$  polarization of R1 and R2 fields, which corresponds to the  angle $\theta =$ 45$^\circ$ or 225$^\circ$ between the optic axis of the QWP and the input polarization (p-polarized) of the R2 field. The Raman fields can not drive the atoms coherently between the states $|1\rangle$ and $|2\rangle$ for polarization combination $(\pi_{1})_x-(\pi_{2})_x$, which corresponds to $\theta =$ 90$^\circ$ or 180$^\circ$. This fact was experimentally confirmed and is presented in Fig.~\ref{fig:s3}(a), which shows that we indeed coherently drive the entire atomic sample as opposed to incoherent driving.  Note that we have observed an additional maximum at $\theta =135^\circ$. The appearance of this maximum can be explained in the above way by considering a $\Lambda$ system with ground states $|F = 2, m_{F} = - 1\rangle \equiv |1\rangle, |F = 2, m_{F} = 0\rangle \equiv |2\rangle$ and excited state $|F' = 3, m_{F'} = -1\rangle \equiv |3\rangle$. Such a combination of states is allowed for the alkali atom $^{87}$Rb in our thermal vapor experiments. 
\par
We here demonstrate the role of angular momentum conservation in coherent coupling through our measurements. In Fig.~\ref{fig:s3}(b), the normalized peak strength of the driven power spectrum ($|\rho_{21}|_{\Delta=0}^2$) from thermal vapors is shown for various angle $\theta$. The observation in Fig.~\ref{fig:s3}(b) shows the fidelity of the coherent coupling of atoms by the Raman fields' polarization state. This can also be applied to control the atomic coherence between ground levels. The manipulation of atomic level coherence may find applications in quantum communications and quantum information processing using neutral atoms.

\begin{figure}[ht]
\centering
\includegraphics[scale=0.36]{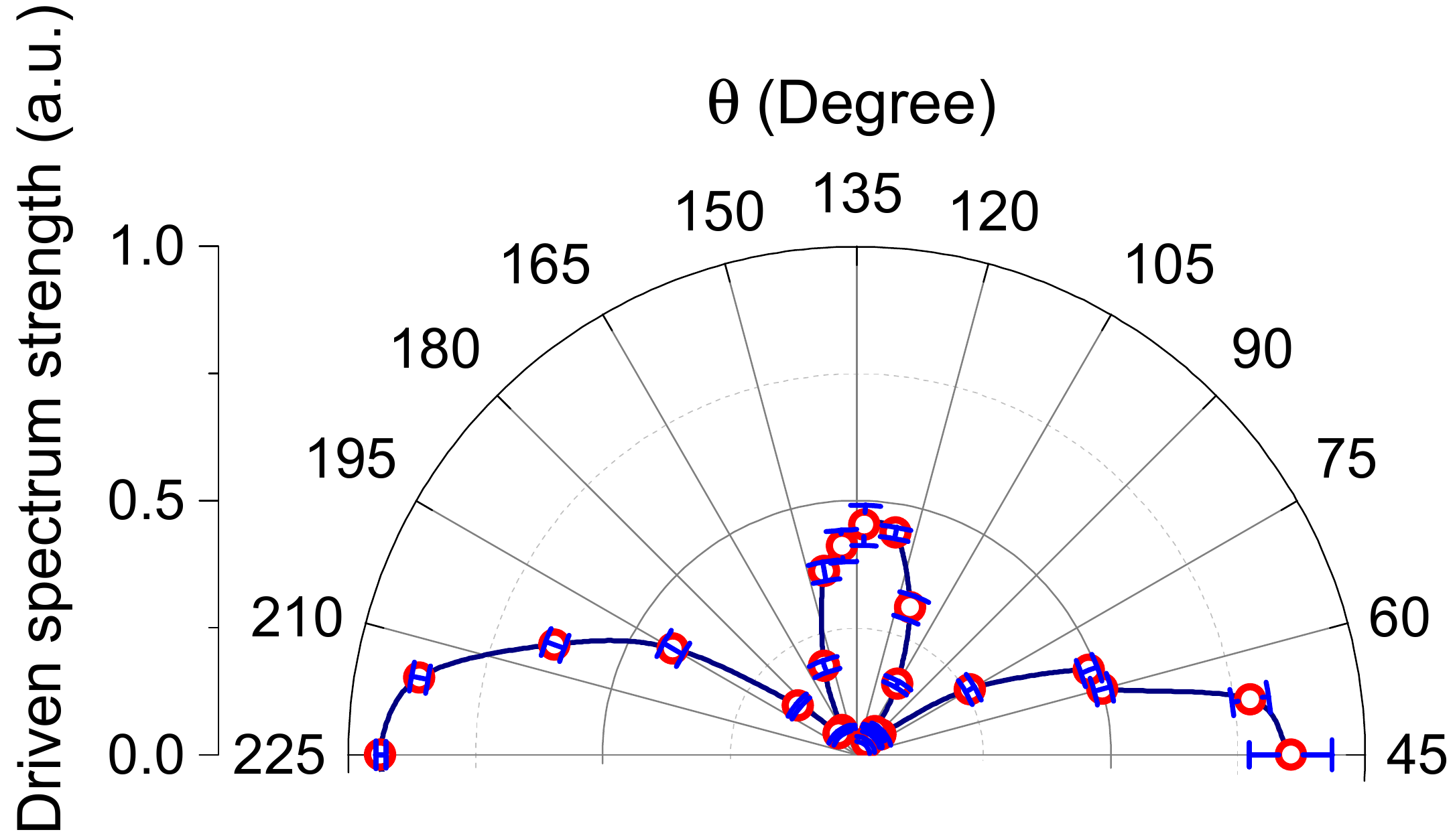}
\renewcommand{\thefigure}{S\arabic{figure}}
\caption{The driven power spectrum strength at $\delta_{12} = 2.73$ MHz for various angle $\theta$ in cold atoms. The signal strength at $\theta = 135^\circ$ is about $1/2$ of that at $\theta = 45^\circ$ and $225^\circ$. The polarization sensitivity at each $\delta_{12}$ carries its unique signature for atoms in the MOT, unlike atoms in a homogeneous magnetic field.}
\label{fig:s4}
\end{figure}

A similar study for the polarization dependence of the driven power spectrum in cold rubidium atoms is shown in Fig.~\ref{fig:s4}. We have fixed $\delta_{12} =$ 2.73 MHz to resonantly drive the atoms located at $x = 1.9$ mm, and $z = 900~\mu$m where $\theta_B \approx 45^\circ$. The dependence of $P(\omega)$ on the Raman fields' polarization for cold atoms is similar to the previous case with thermal vapors (shown in Fig.~\ref{fig:s3}). While the peak signal strength of $P(\omega)$ reduces exactly to zero at $90^\circ$ and $180^\circ$ as expected, the magnitude at $\theta =135^\circ$ reaches only half of the observed values at $\theta=45^\circ$ and $225^\circ$ in cold atoms inside the MOT, which is very different from the thermal vapors in homogeneous magnetic field. 

\subsection{Raman coherence in the quantization basis}

\begin{figure}[h]
\centering
\includegraphics[scale=0.6,trim=0 0 0 0, clip]{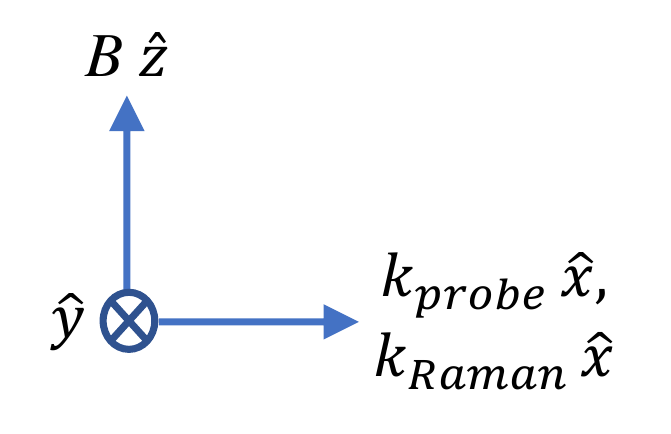}
\renewcommand{\thefigure}{S\arabic{figure}}
\caption{The diagram depicts the direction of the applied uniform magnetic field ($B\hat{z}$), the propagation direction of probe ($k_{probe}\hat{x}$) and Raman ($k_{Raman}\hat{x}$) lasers.}
\label{fig:s5}
\end{figure}

The experiments in vapor cell were performed in the presence of a homogeneous magnetic field (defining the quantization axis) applied along $\hat{z}$, and the Raman fields propagating along $\hat{x}$. The probe laser propagating along $x$-direction detects the $x$-component of the atomic spins. Here we neglect the slight angle between the probe and the Raman lasers, as schematically shown in the Fig.~\ref{fig:s5}.

We consider the polarization state of the Raman fields propagating along $\hat{k}$ $\vert \vert$ $\hat{x}$ being linear ($(\pi_1)_x$) and circular ($(\sigma^{+}_2)_x$). In the presence of the Raman fields, the electronic spins align along the $x$-axis. However, due to the homogeneous magnetic field along $\hat{z}$, the spins precess about $z$-axis on the x-y plane. Since the Larmor precession rate (e.g., $\nu_L \sim$ 4.6 MHz) in our experiments is typically much higher than the spin relaxation rate $(1/2 \pi T_2 \sim 0.15$ MHz), the spins lie on the x-y plane. 

We further restrict our discussion for a system with ground hyperfine level $F = 1$ and excited hyperfine level $F' = 1$. In the following, we will describe the dependence of Raman-driven power spectrum strength on various polarization combinations of R1 and R2 fields presented in Fig.~\ref{fig:s3}. Any spin component of $F$ or $F'$ on the x-y plane can be written as a linear superposition of all possible spin components along $\hat{z}$, for an example \cite{Tekin2016},
\begin{equation}
|m_F = 1\rangle_x = \frac{1}{2}|m_F = -1\rangle_z + \frac{1}{\sqrt{2}}|m_F = 0\rangle_z +\frac{1}{2}|m_F = 1\rangle_z.
\label{Eqn:EngDe}
\end{equation}
We can also decompose the polarization of the Raman fields in terms of their electric fields in the following fashion \cite{Katz2019, Ben2010}: 
\begin{multline}
(\pi_1)_x \equiv \hat{e}_{1y} = \frac{1}{\sqrt{2}}\left( \frac{\hat{e}_{1y} + i \hat{e}_{1x}}{\sqrt{2}}\right) + \frac{1}{\sqrt{2}}\left( \frac{\hat{e}_{1y} - i \hat{e}_{1x}}{\sqrt{2}}\right),\label{Eqn:pi}
\end{multline}
and, 
\begin{eqnarray}
(\sigma_2^{+})_x \equiv \frac{\hat{e}_{2z} + i \hat{e}_{2y}}{\sqrt{2}} &=& \frac{\hat{e}_{2z}}{\sqrt{2}} + \frac{i}{2}\left( \frac{\hat{e}_{2y} + i \hat{e}_{2x}}{\sqrt{2}}\right)\nonumber \\ &+&\frac{i}{2}\left(\frac{\hat{e}_{2y} - i \hat{e}_{2x}}{\sqrt{2}}\right),
\label{Eqn:sigma}
\end{eqnarray}
where, $\hat{e}_{1i}$ or $\hat{e}_{2i}$ is the $i$th component of the corresponding electric field with $i =x,y,z$.

\begin{figure}[h]
\centering
\includegraphics[scale=0.5,trim=0 0 0 0, clip]{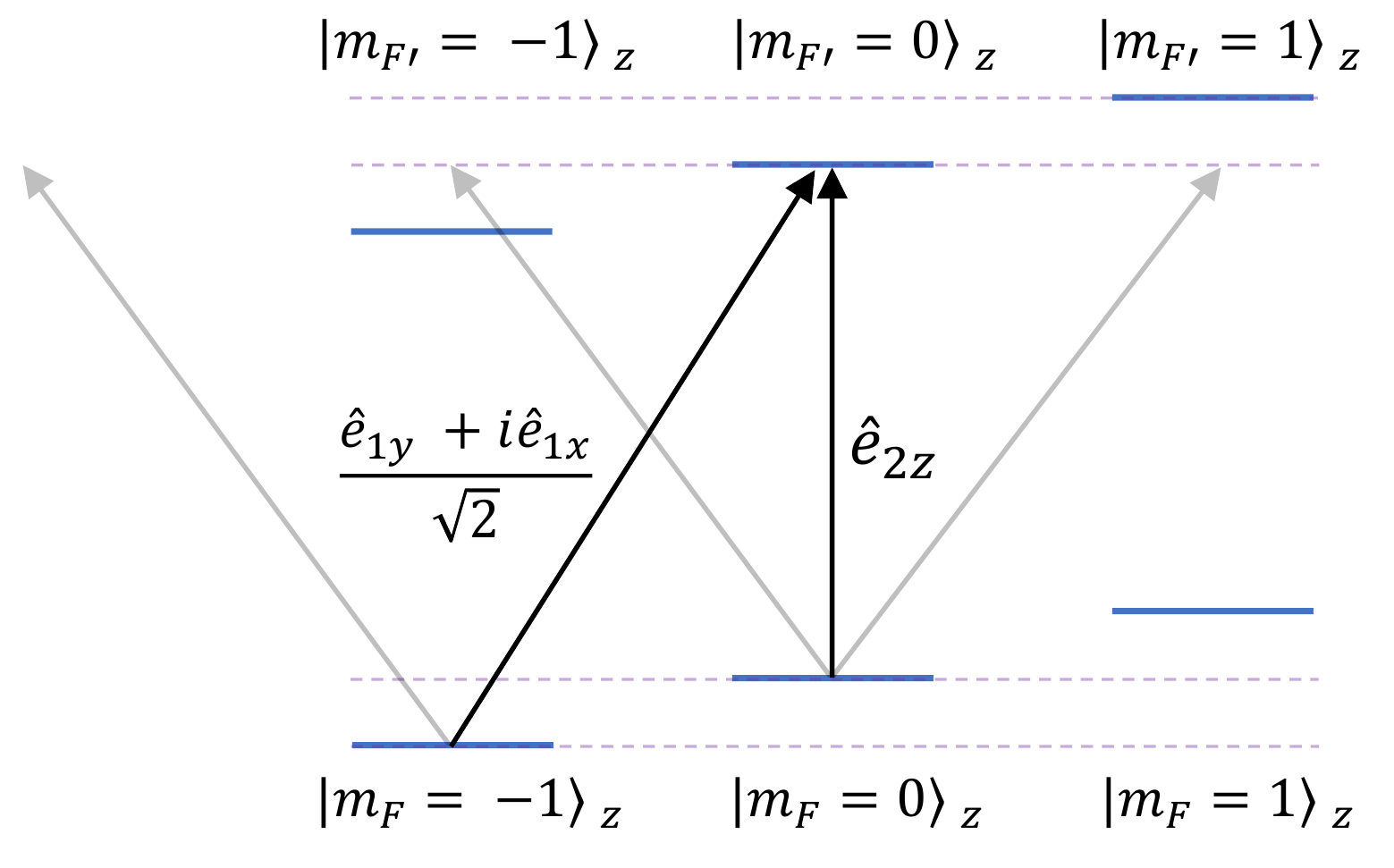}
\renewcommand{\thefigure}{S\arabic{figure}}
\caption{Formation of a $\Lambda$ system in $\hat{z}$ basis within the Zeeman states for $(\pi_1)_x$ and $(\sigma^{+}_2)_x$ combination of polarization of the Raman fields. The generated $\Lambda$ system is shown by black arrows.}
\label{fig:s6}
\end{figure}

In our experiment, we fix the frequency of R1 field (of $(\pi_1)_x$ polarization) on-resonance to $|m_F = -1\rangle_z \leftrightarrow |m_{F'} = 0\rangle_z$ transition, and that of R2 field (of $(\sigma^{+}_2)_x$ polarization) on-resonance to $|m_F = 0\rangle_z \leftrightarrow |m_{F'} = 0\rangle_z$ transition. According to our decomposition in the Eq.~\ref{Eqn:pi} and Eq.~\ref{Eqn:sigma}, the allowed optical transitions in the $\hat{z}$ basis can be shown in Fig.~\ref{fig:s6}, where the field $\hat{e}_{2z}$ couples $|m_F = 0\rangle_z \leftrightarrow |m_{F'} = 0\rangle_z$ transition and the field $(\hat{e}_{1y} + i \hat{e}_{1x})/\sqrt{2}$ couples $|m_F = -1\rangle_z \leftrightarrow |m_{F'} = 0\rangle_z$ transition.

The Fig.~\ref{fig:s6} shows that a $\Lambda$ type 3LS is formed (indicated by black arrows) in $\hat{z}$ basis, and a coherence is built between the states $|m_F = -1\rangle_z \leftrightarrow |m_{F} = 0\rangle_z$. The coherence between $|m_{F} = 0\rangle_z$ and $|m_{F} = 1\rangle_z$ can also be explained in a similar fashion. This coherence in $\hat{z}$ basis in turn built a coherence in $\hat{x}$ basis via Eq.~\ref{Eqn:EngDe}, and detected by the off-resonant probe laser. This case corresponds to $\theta = $ 45$^\circ$ and 225$^\circ$ in Fig.~\ref{fig:s3}. The other maxima at $\theta = $ 135$^\circ$ can be explained by considering the $(\pi_1)_x$ and $(\sigma^{-}_2)_x$ combination of the polarization states of the R1 and R2 field.

For $(\pi_1)_x-(\pi_2)_x$ combination of R1 and R2 field polarizations ($\theta = $ 90$^\circ$ and 180$^\circ$), no $\Lambda$ system is formed in $\hat{z}$ basis within the ground hyperfine level Zeeman states. Therefore, no amplification in the driven power spectrum has been observed.

However, it can be shown using Eq.~\ref{Eqn:sigma} that when both the Raman fields are $\sigma^{+}$ polarized, a double $\Lambda$ system is formed within the consecutive Zeeman states in $F$-manifolds. In this case, the signal strength is two times stronger than the case discussed in Fig.~\ref{fig:s6}.

The Table~\ref{table:pol} summarizes the Raman driven signal strength for various combinations of the Raman fields' polarizations. 
 
\begin{table}[ht]
\centering 
\begin{tabular}{c c} 
\hline\hline
Polarization of Raman fields & Comments on signal strength \\ [0.5ex]
\hline\hline \\ [0.5ex]
$(\pi_1)_x - (\pi_2)_x$ and $(\sigma^{\pm}_1)_x - (\sigma^{\mp}_2)_x$ & No amplification, Intrinsic \\ [1.5ex]
$(\pi_1)_x - (\sigma^{\pm}_2)_x$ and $(\sigma^{\pm}_1)_x - (\pi_2)_x$ & $|\rho_{21}|^2$  \\[1.5ex]
$(\sigma^{\pm}_1)_x - (\sigma^{\pm}_2)_x$ & $2 |\rho_{21}|^2$  \\ [1ex] 
\hline 
\end{tabular}
\caption{Dependence of Raman-driven power spectrum signal strength on the Raman fields' polarization combination.}
\label{table:pol} 
\end{table}

\begin{figure}[h]
\centering
\includegraphics[scale=0.35,trim=0 0 0 0, clip]{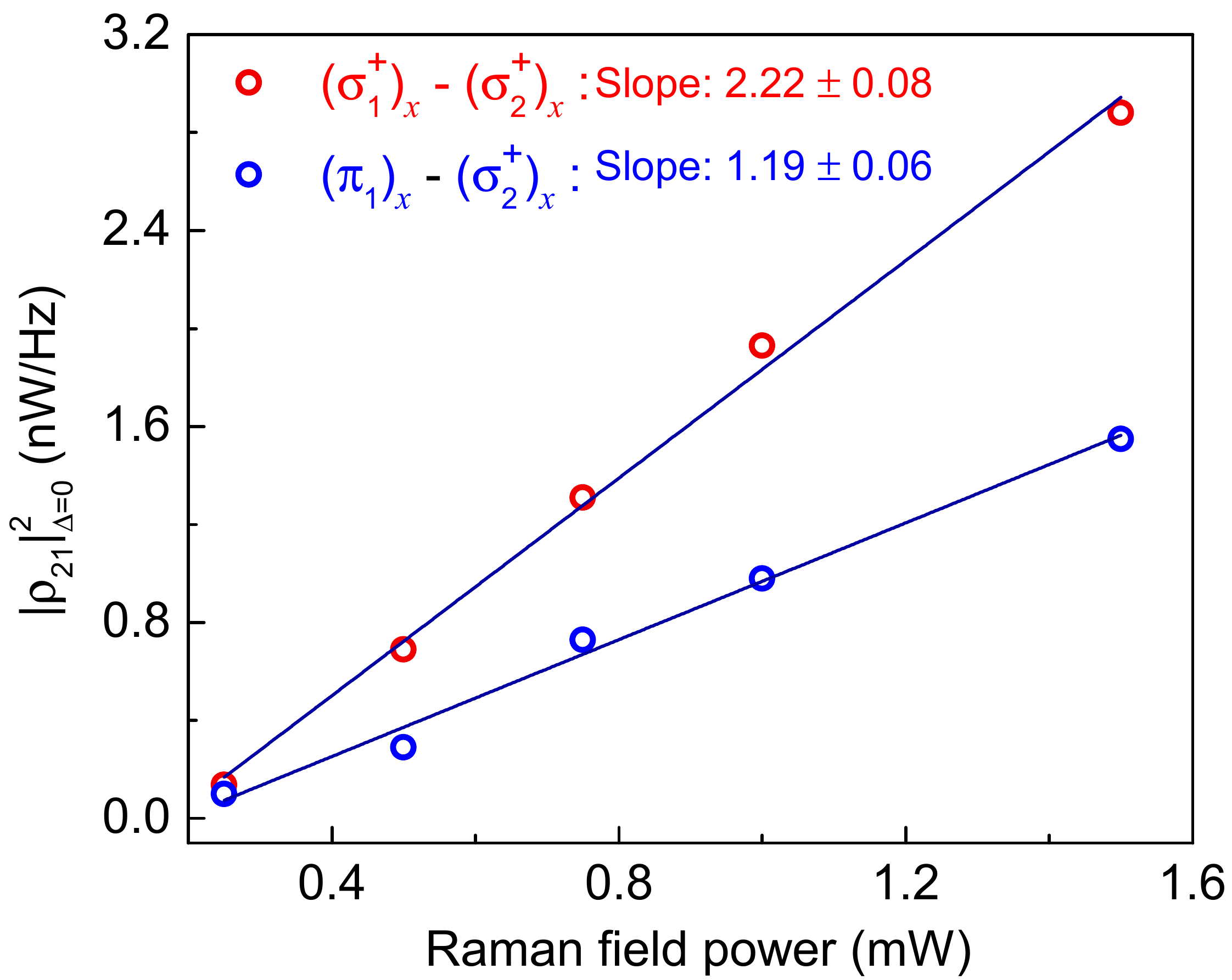}
\renewcommand{\thefigure}{S\arabic{figure}}
\caption{The on-resonance signal strength of the Raman driven spectrum for $(\sigma^{+}_1)_x-(\sigma^{+}_2)_x$ and $(\pi_1)_x -(\sigma^{+}_2)_x$ combination of the Raman fields' polarization with various driving intensities. The plots support the results summarized in Table~\ref{table:pol}.}
\label{fig:s9}
\end{figure}

We have experimentally verified the dependence of the on-resonance Faraday rotation fluctuations signal strength on different polarization combinations of the Raman fields. The experimental results are shown in Fig.~\ref{fig:s9} for completeness.  

\bibliographystyle{apsrev4-1} 
\bibliography{BibSuppli}